\documentclass[onecolumn,12pt]{article}
\usepackage{setspace}
\usepackage{amsmath}
\usepackage{amssymb}
\usepackage{mathtools}
\usepackage{algorithmic}
\usepackage{parskip}
\usepackage[mathscr]{euscript}
\usepackage{graphicx}
\usepackage{caption}
\usepackage{subcaption}
\usepackage{amsthm}
\usepackage{cases}
\usepackage{algorithmic}
\usepackage{parskip}
\usepackage{epstopdf}
\usepackage{color}
\usepackage{makeidx}
\usepackage{natbib}
\usepackage[margin=1in]{geometry}
\newcommand{\rsig}{x(t)}
\newcommand{\tsig}{s(t)}
\newcommand{\nsig}{n(t)}

\newcommand{\mW}{\mathbf{W}}

\newcommand{\mA}{\mathbf{A}}
\newcommand{\mB}{\mathbf{B}}
\newcommand{\mC}{\mathbf{C}}
\newcommand{\mSS}{\mathbf{D}}
\newcommand{\mBB}{\tilde{\mathbf{B}}}

\newcommand{\mytop}{T}
\newcommand{\va}{\mathbf{a}}
\newcommand{\vx}{\mathbf{x}}
\newcommand{\vb}{\mathbf{b}}

\newcommand{\vvv}{\tilde{\mathbf{v}}}
\newcommand{\pfa}{P_{\mbox{\scriptsize FA}}}
\newcommand{\pdet}{P_{\mbox{\scriptsize D}}}

\newcommand{\myL}{L}

\newcommand{\vv}{\mathbf{v}}
\newcommand{\vn}{\mathbf{n}}

\newcommand{\vh}{\mathbf{h}}
\newcommand{\vc}{\mathbf{c}}
\newcommand{\Nef}{N^\prime}
\newcommand{\ww}{\mathbf{w}}

\newcommand{\rrz}{\tilde{\mathbf{r}}_z}
\newcommand{\vech}{\mathrm{vech}}
\newcommand{\vect}{\mathrm{vec}}

\newcommand{\xrl}{\acute{x}}

\newcommand{\xiy}{\grave{x}}

\newcommand{\myyvskip}{\vskip-6pt}
\newcommand{\myyhskip}{\hskip+10pt}
\newcommand{\myvhskip}{\vskip-6pt \hskip+10pt}

\newcommand{\Td}{{T}}
\newcommand{\Tcp}{T_{cp}}
\newcommand{\Tsym}{T_{sym}}
\newcommand{\wij}{w_{i,j}}
\newcommand{\wpq}{w_{p,q}}
\newcommand{\rij}{r_{i,j}}
\newcommand{\rpq}{r_{p,q}}
\newcommand{\cH}{\mathcal{H}}

\newcommand{\Rh}{\bar{\mR}}
\newcommand{\rhz}{\bar{\vr}_z}
\newcommand{\mLambda}{\mbox{\boldmath$\Lambda$\unboldmath}}
\newcommand{\mGamma}{\mbox{\boldmath$\Gamma$\unboldmath}}
\newcommand{\mSigma}{\mbox{\boldmath$\Sigma$\unboldmath}}
\newcommand{\mDelta}{\mbox{\boldmath$\Delta$\unboldmath}}
\newcommand{\Hmat}{\mbox{\boldmath$\Omega$\unboldmath}}
\newcommand{\Gmat}{\mbox{\boldmath$\Gamma$\unboldmath}}
\newcommand{\thetab}{\mbox{\boldmath$\theta$\unboldmath}}

\newcommand{\Nblk}{{N_B}}

\newcommand{\Tone}{T_1}
\newcommand{\Ttwo}{T_2}

\newcommand{\nonr}{\nonumber}
  
\newcommand{\beq}{\begin{equation}} 
\newcommand{\eeq}{\end{equation}} 
\newcommand{\beqa}{\begin{eqnarray}}
\newcommand{\eeqa}{\end{eqnarray}}

\newcommand{\vsig}{\mathbf{x}}  
\newcommand{\vsnsig}{\mathbf{z}}
\newcommand{\smpMat}{\mathbf{A}}

\newcommand{\mR}{\mathbf{R}}

\newcommand{\mS}{\mathbf{S}}

\newcommand{\mJ}{\mathbf{J}}

\newcommand{\Exp}{\mathbb{E}}

\newcommand{\cN}{\mathcal{N}}

\newcommand{\cC}{\mathcal{C}}

\newcommand{\vy}{\mathbf{y}}
\newcommand{\vr}{\mathbf{r}}
\newcommand{\vs}{\mathbf{s}}

\newcommand{\mI}{\mathbf{I}}
\newcommand{\mK}{\mathbf{K}}
\newcommand{\mZero}{\mathbf{0}}

\newtheorem{thm}{Theorem}%[section]

\begin{document}

%\begin{frontmatter}



%\title{}


%\author{}

%\address{}

% Example definitions.
% --------------------
%\def\x{{\mathbf x}}
%\def\L{{\cal L}}
%\newcommand{\sK}{\mathscr{K}}
% Title.
% ------
\title{Covariance-Based OFDM Spectrum Sensing with Sub-Nyquist Samples}
%
% Single address.
% ---------------
%\author{A. Author-one, B. Author-two, C. Author-three}% \\ Department of Electronics and Communications Engineering, \\ Tampere University of Technology, Tampere, Finland}%\thanks{Thanks to XYZ agency 

%\author[tut]{S. Alireza Razavi\corref{fn1}}%\corref{cor1}\fnref{fn1}}
%\ead{alireza.razavi@tut.fi}
%\author[tut]{Mikko Valkama}%\fnref{fn2}}
%\ead{mikko.e.valkama@tut.fi}
%\author[ucla]{Danijela Cabric}%\corref{cor2}\fnref{fn1,fn3}}
%\ead{danijela@ee.ucla.edu}
%%\ead[url]{http://www.elsevier.com}
%\cortext[fn1]{Corresponding author.}
%%\fntext[cor1]{The work of A. Razavi and M. Valkama was supported by the Academy of Finland under the  project $\#$251138 ``Digitally-Enhanced RF for Cognitive Radio Devices'' and the Finnish Funding Agency for Technology and Innovation (Tekes), under the project ``Enabling Methods for Dynamic Spectrum Access and Cognitive Radio (ENCOR)''.}

%\address[tut]{Department of Electronics and Communications Engineering, Tampere University of Technology, Tampere, Finland.}
%\address[ucla]{Cognitive Reconfigurable Embedded Systems Lab (CORES), University of California, Los Angeles (UCLA).}
\author{Alireza Razavi$^\dagger$, Mikko Valkama$^\dagger$, Danijela Cabric$^\ddagger$ \\ 
$^\dagger$Tampere University of Technology, Tampere, Finland\\
$^\ddagger$University of California, Los Angeles (UCLA), USA \\ 
email: alireza.razavi@tut.fi, mikko.e.valkama@tut.fi, danijela@ee.ucla.edu
}

%\author{Alireza Razavi~~~~~~%,~\IEEEmembership{Member,~IEEE,}
        %Mikko Valkama~~~~~~%,~\IEEEmembership{Member,~IEEE,}
        %Danijela Cabric}%,~\IEEEmembership{Member,~IEEE}% <-this % stops a space
%\address{A. Razavi and M. Valkama are with the Department
%of Electronics and Communications Engineering, Tampere University of Technology, Tampere, Finland. D. Cabric is with the Cognitive Reconfigurable Embedded Systems Lab (CORES) University of California, Los Angeles (UCLA).}% <-this % 
%\thanks{This work was supported by the Academy of Finland under the  project $\#$251138 ``Digitally-Enhanced RF for Cognitive Radio Devices'' and the Finnish Funding Agency for Technology and Innovation (Tekes), under the project ``Enabling Methods for Dynamic Spectrum Access and Cognitive Radio (ENCOR)''.}}

%\begin{document}

\date{}
%

%\ninept
%
\maketitle
%\makeindex
%
\begin{abstract}
In this paper, we propose a feature-based method for spectrum sensing of OFDM signals from sub-Nyquist samples over a single band. 
We exploit the structure of the covariance matrix of OFDM signals to convert an underdetermined set of covariance-based equations to an overdetermined one.
The statistical properties of sample covariance matrix are analyzed and then based on that an approximate Generalized Likelihood Ratio Test (GLRT) for detection of OFDM signals from sub-Nyquist samples is derived. The method is also extended to the frequency-selective channels.
\end{abstract}
%
%\begin{keywords}
%Carrier Frequency Offset, Sub-Nyquist Sampling, Cyclostationary
%\end{keywords}
%
%\begin{keyword}
%Spectrum sensing, OFDM, sub-Nyquist sampling, compressive sensing, cognitive radio.
%\end{keyword}
%\end{frontmatter}
%%%%%%%%%%%%%%%%%%%%%%%%%%%%%%%%%%%%%%%%%%%%%%%%%%%%%%%%%%%%%%%%%%%%%%%%%%%%%%%%%%%%%%%%%%%%%%%%%%%%%%%%%%%%%%%%%%%%%%%%%%%%%%%%%%%%%%%%%%%%%%%%%%%%%%%%%%%%%%%%%%%%%%%%%%%%%%%%%%%%%%%%%%%%%%%%
%%%%%%%%%%%%%%%%%%%%%%%%%%%%%%%%%%%%%%%%%%%%%%%%%%%%%%%%%%%%%%%%%%%%%%%              SECTION I            %%%%%%%%%%%%%%%%%%%%%%%%%%%%%%%%%%%%%%%%%%%%%%%%%%%%%%%%%%%%%%%%%%%%%%%%%%%%%%%%%%%%%%
%%%%%%%%%%%%%%%%%%%%%%%%%%%%%%%%%%%%%%%%%%%%%%%%%%%%%%%%%%%%%%%%%%%%%%%%%%%%%%%%%%%%%%%%%%%%%%%%%%%%%%%%%%%%%%%%%%%%%%%%%%%%%%%%%%%%%%%%%%%%%%%%%%%%%%%%%%%%%%%%%%%%%%%%%%%%%%%%%%%%%%%%%%%%%%%%
\section{Introduction}
\label{sec:intro}
\myyhskip Cognitive Radio (CR) is emerging as a promising technology for improving the efficiency of radio spectrum use in wireless communication systems \citep{cr99}.
%The main task in CR systems is Spectrum Sensing (SS) which aims in detecting the spectral holes, i.e. licensed frequency bounds which has not been utilized by the primary users (PU),  and allocating them to 
%the secondary users (SU). 
%Cognitive radio emerged to response the ever-increasing demand for higher data rates and new wireless services by solving the spectrum underutilization problem \cite{cr99, haykin05}.
Spectrum sensing (SS) is the most vital task in CR defined as identifying spectrum holes by sensing the radio spectrum and utilizing them
without causing interference to primary users (PUs) \citep{haykin09}. 
Of special interest in this regard is sensing of OFDM signals \citep{sachin09, axell11}. OFDM is one of the most effective multicarrier techniques for broadband wireless communications which is employed by
many of the current and emerging wireless technologies.%, such as WiFi, WiMAX, LTE, etc. 

%In this context, a broader spectral awareness could potentially detect and exploit more spectral underutilizations and 
%therefore increase the capacity. 
\myvhskip On the other hand, due to the limitations of today's analog-to-digital converter (ADC) circuits which cannot support very high bandwidth and need excessive memory and prohibitive
energy costs for implementing digital signal processing systems \citep{cohen11}, it may be very costly and even impractical to sense the signal based on Nyquist-rate
samples. 
This has motivated researchers to study sub-Nyquist methods for wideband spectrum sensing in CR networks; see, e.g., \citep{mishali11, tian07, tian12, polo09, tian_icc11, leus11, ariananda11, cohen11, sun12, rebeiz12}. But to the best of our knowledge,
there has been less efforts targeting the detection of OFDM signals from sub-Nyquist samples. In this paper, we propose a new feature-based approach for sensing an OFDM signal occupying a single band from sub-Nyquist samples.

{\bf Related research and previous work:} The problem of OFDM sensing using second-order statistics has been already studied in, e.g., \citep{sachin09, axell11, simin11, dobre12}. All of these methods in some way exploit the correlation induced by CP to sense the presence of OFDM signal, but perform the detection based on Nyquist rate samples which, as discussed earlier, might need expensive ADCs espeially for wideband signals. 

%For sub-Nyquist sampling of a signal, There are two main strategies proposed in the literature \cite{xampling}: Random Demodulator (RD), also known as Analog-to-Information Converter (AIC), \cite{tropp10, kirolos06}, and Modulated Wideband Converter (MWC) \cite{mishali10}. While MWC is suitable for multiband signals, RD can be employed also for sub-Nyquist sampling of single-band signals. The proposed method in this paper employs the RD strategy for sub-Nyquist sampling of OFDM signals as its goal is to decide on occupancy of a single band. 

During the past several years, the problem of spectrum sensing from sub-Nyquist samples has attracted a lot of attention. While many of the approaches proposed so far rely on the sparsity in the frequency domain arising from spectrum underutilization \cite{tian07, tian08, polo09, sun13}, there has also been methods which do not necessarily need sparseness in the spectrum to work \cite{tian12, tian_icc11, leus11, ariananda11, razavi13, cohen11, rebeiz12}. 
%Employing RD for compressive spectrum sensing has been already addressed in several papers \cite{tian12, tian_icc11, polo09, leus11, ariananda11, razavi13, cohen11, rebeiz12}. 
For instance, the methods proposed in \cite{tian12, tian_icc11, leus11, cohen11, rebeiz12, razavi13} are based on  recovering the Spectral Correlation Function (SCF) of the signal from sub-Nyquist samples.  All of these methods need sparsity in the SCF for detecting the signal. Since the SCF matrix of the Nyquist-sampled OFDM signal (see, e.g., \cite[Equation (10)]{tian12}) is nonzero everywhere, then it is not possible to use these methods for recovering the SCF of OFDM signals from sub-Nyquist samples as they need sparsity in the SCF for recovering it using Compressive Sensing (CS) algorithms or converting the underdetermined set of equations to an overdetermined one. The methods proposed in \cite{ariananda11} and \cite[Section IV]{tian12} reconstruct the Power Spectral Density (PSD) from sub-Nyquist samples and then decide on the presence of signal based on the recovered PSD. Similar to conventional energy detection methods \cite{yucek09}, the main challenges with these methods are selecting the threshold and poor performance in low signal-to-noise ratios.  In \cite{razavi13}, we detected an OFDM signal over a single band from sub-Nyquist samples. Although this method assume that the whole band is occupied by the OFDM signal, but we inject the sparsity to SCF using a cyclostationary signature \cite{adrat09, sutton08} embedded in the signal to assist with the detection procedure. This might be of help in the problem of rendezvous and cognitive network identification \cite{razavi14}, but it cannot be used in detecting primary users, as primary users usually do not care about secondary users and therefore do not assist them via injecting a signature in their own signals to facilitate the spectrum sensing task. 

{\bf Our contribution:} In this paper, we propose a new feature-based approach for sensing OFDM signals from sub-Nyquist samples over a single band entirely occupied by the signal. In other words, we do not assume any sparsity in the spectrum. Instead, the method exploits the features of covariance matrix of OFDM signals stemming from the insertion of cyclic prefix (CP). These features, as we will show later, will help us to convert the underdetermined set of two-dimensional (2-D) equations which relates the available covariance matrix of sub-Nyquist samples to the unavailable covariance matrix of Nyquist samples, to an overdetermined one-dimensional (1-D) one. Then, based on the statistical properties of the sample covariance, an approximate GLRT-based detector is derived. The method is also extended to the case of frequency-selective channels.

{\bf Paper organization:} The rest of the paper is organized as follows. In Section \ref{sec:2} the system model over Gaussian channels is given and the problem is formulated. In Section \ref{sec:3} the relevant statistical properties of sample covariance matrix are studied and a covariance-domain linear system is derived. The approximate GLRT-based detector is introduced and formulated in Section \ref{sec:4}. In Section \ref{sec:multipath} we extend the results to the frequency-selective channel case. Finally, we study the performance of the proposed method by simulation experiments in Section \ref{sec:5}. Conclusions are drawn in Section \ref{sec:conclusion} and some details of the derivations are given in the Appendix. 

{\bf Notations and Mathematical Preliminaries:} Throughout this paper matrices and vectors are denoted by capital and small boldface letters, respectively. $=$ denotes the equality and $\triangleq$ denotes the definition.
$\Exp$ is reserved for statistical expectation operator, $\mbox{Cov}(\vx,\vy)$ represents the covariance matrix between random vectors $\vx$ and $\vy$, and $\otimes$ denotes the Kronecker product.
$\mI_P$ and $\mZero_{P,Q}$ represent, respectively,
$P \times P$ identity matrix and $P \times Q$ full-zero matrix. For an arbitrary $M \times N$ matrix $\smpMat$, $[\smpMat]_{i,j}$ denotes the $(i,j)$-th entry of the matrix and $\va_i,~i=1,2,\ldots,N$ denotes its $i$-th column. $\vect(\mA)$ is the vectorization of $\mA$ defined as $\vect(\mA)\triangleq[\va_1^T,\ldots,\va_N^T]^T$. Furthermore, if $M=N$, then  $\vech(\mA)$, known as half-vectorization of $\mA$, is the $\frac{N(N+1)}{2} \times 1$ vector which is obtained by column-wise stacking of only the elements on and below the main diagonal. It can be proven that for an $N \times N$ matrix $\mA$ \citep[Chapter 11]{seber08}
\beq
\vech(\mA)= \Hmat_N \vect(\mA),
\label{eq:vech}
\eeq 
where  
\beqa
\Hmat_N \triangleq \left[\begin{array}{ccccc}
\mI_N^{(0)} & \mZero_{N,N} & \mZero_{N,N} & \ldots & \mZero_{N,N} \\
\mZero_{N-1,N} & \mI_N^{(1)} & \mZero_{N-1,N} & \ldots & \mZero_{N-1,N} \\
\mZero_{N-2,N} & \mZero_{N-1,N} & \mI_N^{(2)} & \ldots & \mZero_{N-2,N} \\
\vdots & \vdots & \vdots & \ddots & \vdots \\
\mZero_{1,N} & \mZero_{1,N} & \mZero_{1,N}  & \ldots & \mI_N^{(N-1)} \\
\end{array}\right],
\label{eq:Hmat}
\eeqa
where $\mI_N^{(i)}$ denotes an identity matrix whose first $i$ rows are discarded. Remark that $\mI_N^{(0)} = \mI_N$. Moreover, if $\mA$ is a symmetric matrix, then \citep[Chapter 11]{seber08}
\beq
\vect(\mA)= \Gmat_N \vech(\mA),
\label{eq:vect}
\eeq 
where $\Gmat$ is an $(N^2)\times(N(N+1)/2)$ matrix with entries $((i-1)N+j,(j-1)(N-j/2)+i)$ and $((j-1)N+i,(j-1)(N-j/2)+i)$ for $1 \le j \le i \le N$ equal to $1$ and the rest of elements equal to zero. Remark that $\Hmat_N \Gmat_N=\mI_{N(N+1)/2}$.

Furthermore, for any three arbitrary matrices $\mA$, $\mB$, and $\mC$ of suitable sizes, we have the following two equalities \cite[Chapter 11]{seber08}
%\begin{itemize}	\item 
\beq
\vect(\mA \mB \mC^T) = (\mC \otimes \mA) \vect(\mB).
\label{eq:veckron}
\eeq
and
%\item 
\beq
\mA \mB \mC^T = \sum_i \sum_j [\mB]_{i,j} \va_i \vc_j^T,
\label{eq:veckron2}
\eeq
where $\va_i$ and $\vc_j$ are $i$-th and $j$-th columns of $\mA$ and $\mC$, respectively. 
%\end{itemize}
%%%%%%%%%%%%%%%%%%%%%%%%%%%%%%%%%%%%%%%%%%%%%%%%%%%%%%%%%%%%%%%%%%%%%%%%%%%%%%%%%%%%%%%%%%%%%%%%%%%%%%%%%%%%%%%%%%%%%%%%%%%%%%%%%%%%%%%%%%%%%%%%%%%%%%%%%%%%%%%%%%%%%%%%%%%%%%%%%%%%%%%%%%%%%%%%
%%%%%%%%%%%%%%%%%%%%%%%%%%%%%%%%%%%%%%%%%%%%%%%%%%%%%%%%%%%%%%%%%%%%%%%             SECTION II            %%%%%%%%%%%%%%%%%%%%%%%%%%%%%%%%%%%%%%%%%%%%%%%%%%%%%%%%%%%%%%%%%%%%%%%%%%%%%%%%%%%%%%
%%%%%%%%%%%%%%%%%%%%%%%%%%%%%%%%%%%%%%%%%%%%%%%%%%%%%%%%%%%%%%%%%%%%%%%%%%%%%%%%%%%%%%%%%%%%%%%%%%%%%%%%%%%%%%%%%%%%%%%%%%%%%%%%%%%%%%%%%%%%%%%%%%%%%%%%%%%%%%%%%%%%%%%%%%%%%%%%%%%%%%%%%%%%%%%%
\section{System Model over Gaussian Channels}
\label{sec:2}

\myvhskip Consider a secondary user (SU) with the goal to detect the presence of an OFDM signal.
Denoting the received signal by $\rsig$, we can formulate the problem as deciding between the following two hypotheses
\beqa
\left\{ \begin{array}{l} \cH_0: \rsig = \nsig \\ 
\cH_1: \rsig = \tsig + \nsig \end{array} \right.
\label{eq:test1}
\eeqa
where $\tsig$ is an OFDM signal and $\nsig \sim \cC \cN(0,\sigma_n^2) $ . If the number of subcarriers is large enough, from central limit theory we have
$\tsig \sim \cC \cN(0,\sigma_s^2)$. Suppose next that we have sampled the signal $\rsig$ at a sub-Nyquist rate to collect
compressive samples $z(t)$. Matrix-wise, this can be described as\footnote{We remark here that there are two main strategies for sub-Nyquist sampling of a signal \cite{xampling}: Random Demodulator (RD), also known as Analog-to-Information Converter (AIC), \cite{tropp10, kirolos06}, and Modulated Wideband Converter (MWC) \cite{mishali10}. Both of these strategies can be mathematically formulated as in (\ref{subnyqsamp}) \cite{xampling}.}
\beq
\vsnsig(k) = \smpMat \vsig(k),~k=1,2,\ldots,\Nblk,
\label{subnyqsamp}
\eeq
where $\vsnsig(k)\triangleq[z[kM],z[kM+1],\ldots,z[kM+M-1]]^\mytop$ consists of sub-Nyquist (compressive) samples, $\vsig(k)\triangleq[x[kN],x[kN+1],\ldots,x[kN+N-1]]^\mytop$ consists of (unavailable) Nyquist samples, $\Nblk$ is the total number of taken frames,
and $\smpMat$ is the $M \times N$ real-valued measurement matrix. $\rho \triangleq \frac{M}{N}<1$ is called compression (or downsampling) ratio. 

Now, the task of a sub-Nyquist OFDM detector is to 
decide whether the OFDM signal exists or not, based on sub-Nyquist measurements $\{\vsnsig(k)\}_{k=1}^\Nblk$.
%In this paper we will show how this can be done using the properties of the autocorrelation (AC) of OFDM signal.

\myvhskip It can be easily seen that $\mR_z=\smpMat \mR_x \smpMat^T$
where the $N \times N$ matrix $\mR_x\triangleq \Exp(\vsig(k) \vsig^H(k))$ and $M \times M$ matrix $\mR_z\triangleq \Exp(\vsnsig(k) \vsnsig^H(k))$ are covariance matrices of $\vsig$ and $\vsnsig$, respectively. 
Let us denote the useful symbol length of the considered OFDM signal by $\Td$, the cyclic prefix length by $\Tcp$ and the total OFDM symbol length by $\Tsym=\Td+\Tcp$ which are all assumed to be known to the cognitive user.
Assume that the basic sample duration is normalized to 1. Then by setting $N=\Tsym$, matrix $\mR_x$ of the noisy OFDM signal (i.e. $x(t)$ under $\cH_1$) can be written as
  
\beqa
[\mR_x]_{i,j}=\left\{
\begin{array}{ll}
\sigma_s^2+\sigma_n^2 & \mbox{if}~i=j,\\
\sigma_s^2 & \mbox{if}~|i-j|=\Td,\\
0 & \mbox{otherwise}.
\end{array}\right.
\label{eq:rh1}
\eeqa
On the other hand, under $\cH_0$ we have
\beqa
[\mR_x]_{i,j}=\left\{
\begin{array}{ll}
\sigma_n^2 & \mbox{if}~i=j,\\
0 & \mbox{otherwise}.
\end{array}\right.
\eeqa
\myvhskip To encompass the descriptions of $\mR_x$ under $\cH_0$ and $\cH_1$, we present it by 
\beq
\mR_x = \tau_0 \mI_N + \tau_s \mLambda,
\label{eq:Rx_OFDM}
\eeq
where $\mI_N$ is the $N \times N$ identity matrix and $\mLambda$ is defined as
\beqa
[\mLambda]_{i,j} \triangleq \left\{ \begin{array}{ll}
1 & \textrm{if}~|i-j|=T, \\
0 & \textrm{otherwise},
\end{array} \right.
\eeqa
and then distinguish between $\cH_0$ and $\cH_1$ as
\beqa
\left\{ \begin{array}{l} \cH_0: (\tau_0,\tau_s) = (\sigma_n^2,0), \\ 
\cH_1: (\tau_0,\tau_s) = (\sigma_n^2+\sigma_s^2,\sigma_s^2). \end{array} \right.
\label{eq:test11}
\eeqa
\myvhskip Since in both cases of (\ref{eq:test11}) we have $\tau_0=\tau_s+\sigma_n^2$, therefore (\ref{eq:test11}) can be simplified as
\beqa
\left\{ \begin{array}{l} \cH_0: \tau_s = 0, \\ 
\cH_1: \tau_s \neq 0 . \end{array} \right.
\label{eq:test111}
\eeqa
In other words, the problem of detection of OFDM signal can be re-expressed as identifying whether in the general description of $\mR_x$ in (\ref{eq:Rx_OFDM}) the 
parameter $\tau_s$ is zero or not. 

\myvhskip In practice, covariance matrices are not readily available and hence we substitute them by sample covariance matrices $\Rh_x\triangleq\frac{1}{\Nblk} \sum_{k=1}^\Nblk \vsig(k) \vsig^H(k)$ and 
$\Rh_z=\frac{1}{\Nblk} \sum_{k=1}^\Nblk \vsnsig(k) \vsnsig^H(k)$, where $\Nblk$ is the number of frames used for the computations.
It is easy to see that
\beq
\Rh_z=\smpMat \Rh_x \smpMat^\mytop.
\label{hRz_vs_hRx}
\eeq
In the next sections we will show how the sub-Nyquist sample covariance matrix $\Rh_z$ can be used to perform the hypothesis testing in (\ref{eq:test1}) based on the observation made in (\ref{eq:test111}).

%%%%%%%%%%%%%%%%%%%%%%%%%%%%%%%%%%%%%%%%%%%%%%%%%%%%%%%%%%%%%%%%%%%%%%%%%%%%%%%%%%%%%%%%%%%%%%%%%%%%%%%%%%%%%%%%%%%%%%%%%%%%%%%%%%%%%%%%%%%%%%%%%%%%%%%%%%%%%%%%%%%%%%%%%%%%%%%%%%%%%%%%%%%%%%%%
%%%%%%%%%%%%%%%%%%%%%%%%%%%%%%%%%%%%%%%%%%%%%%%%%%%%%%%%%%%%%%%%%%%%%%%             SECTION III           %%%%%%%%%%%%%%%%%%%%%%%%%%%%%%%%%%%%%%%%%%%%%%%%%%%%%%%%%%%%%%%%%%%%%%%%%%%%%%%%%%%%%%
%%%%%%%%%%%%%%%%%%%%%%%%%%%%%%%%%%%%%%%%%%%%%%%%%%%%%%%%%%%%%%%%%%%%%%%%%%%%%%%%%%%%%%%%%%%%%%%%%%%%%%%%%%%%%%%%%%%%%%%%%%%%%%%%%%%%%%%%%%%%%%%%%%%%%%%%%%%%%%%%%%%%%%%%%%%%%%%%%%%%%%%%%%%%%%%%
\section{Covariance-Based Linear Equations for OFDM Detection}
\label{sec:3}
\myvhskip The sample covariance matrix $\Rh_x$ can in general be written as the sum of covariance matrix $\mR_x$ and an error term $\mW$ stemming from the finite-sample effects. From now on, we call $\mW$ {\it the finite-sample noise} and express this as
\beq
\Rh_x = \mR_x + \mW.
\label{eq:Rx_noisy}
\eeq
It is easy to verify that $\Exp_x(\Rh_x)=\mR_x$ under both $\cH_0$ and $\cH_1$, and therefore conclude that $\Exp_x(\mW)=\mZero$.
Furthermore, from \citep[Chapters~9-10]{goldberger91} we know that the entries of $\mW$ have asymptotic Normal distribution.
The following theorem then states the covariance of entries of $\mW$. Parts of this theorem can also be found in \cite{axell11}.
\begin{thm}
\label{thm1}
Suppose that $w_{i,j}$ and $w_{p,q}$ are two arbitrary entries of $\mW$ below or on main diagonal (i.e. $i \ge j$, $p \ge q$). Then 
\beq
\Exp(\wij\wpq|\cH_0)=\left\{\begin{array}{ll}
\frac{\tau_0^2}{2\Nblk} & \textrm{if}~(i,j)=(p,q) \\
0 & \textrm{otherwise},
\end{array}
\right.
\label{eq:covar_h0}
\eeq
where $\tau_0 \triangleq \sigma_n^2$, and 
%\begin{numcases}{\Exp(\wij \wpq|\cH_1)=}
%\frac{\tau_0^2}{L},              & for $(i,j)=(p,q),i-j=0$ \nonr \\
%\frac{\tau_s^2+\tau_0^2}{2L},    & for $(i,j)=(p,q),i-j=\Td$ \nonr\\
%\frac{\tau_0^2}{2L},             & for $(i,j)=(p,q),i-j \notin \{0,T\}$ \nonr\\
%\frac{\tau_0 \tau_s}{L},         & for $i=j=p=q+\Td$ \nonr\\
%0.                               & otherwise
%\end{numcases}
\beqa
\Exp(\wij \wpq|\cH_1)=\begin{cases}
\frac{\tau_0^2}{\Nblk} & \textrm{if}~(i,j)=(p,q),i-j=0,\\
\frac{\tau_s^2+\tau_0^2}{2\Nblk} & \textrm{if}~(i,j)=(p,q),i-j=\Td,\\
\frac{\tau_0^2}{2\Nblk} & \textrm{if}~(i,j)=(p,q),i-j \notin \{0,T\}\\
\frac{\tau_0 \tau_s}{\Nblk} & \textrm{if}~~ i=j=p=q+\Td \\
0 & \textrm{otherwise}, \end{cases}  \label{eq:var_h1} \eeqa
%0 & \textrm{otherwise}. ~~~~~~~~~~~~~~~~~~~~(\arabic{equation})\label{eq:var_h1}\end{cases} \nonr \eeqa
%\addtocounter{equation}{1}
%\end{array}
%\beqa
%\Exp(\wij \wpq|\cH_1)=\left\{\begin{array}{ll}
%\frac{\tau_0^2}{L} & \textrm{if}~(i,j)=(p,q),i-j=0,\\[.2em]
%\frac{\tau_s^2+\tau_0^2}{2L} & \textrm{if}~(i,j)=(p,q),i-j=\Td,\\[.2em]
%\frac{\tau_0^2}{2L} & \textrm{if}~(i,j)=(p,q),i-j \notin \{0,T\}\\[.2em]
%\frac{\tau_0 \tau_s}{L} & \textrm{if}~~ i=j=p=q+\Td \\[.2em]
%0 & \textrm{otherwise}.
%\end{array}
%\right. \nonr \eeqa
%\beq
%\label{eq:var_h1}
%\eeq 
%\end{itemize}
\end{thm}
where $\tau_0 \triangleq \sigma_n^2+\sigma_s^2$ and $\tau_s \triangleq \sigma_s^2$.
\myyvskip \myyvskip\begin{proof}
Proof is deferred to Appendix.
\end{proof}

\myvhskip Inserting next (\ref{eq:Rx_noisy}) in (\ref{hRz_vs_hRx}) yields
\beq
\Rh_z = \smpMat \mR_x \smpMat^\mytop + \smpMat \mW \smpMat^\mytop.
\label{eq:hRz_vs_Rx}
\eeq
Notice that since $\mR_x$ and $\smpMat$ are both real-valued, the signal part of (\ref{eq:hRz_vs_Rx}) is real-valued and the imaginary part only includes
the effect of finite-sample error $\mW$. 
In fact, we can simply change (\ref{eq:hRz_vs_Rx}) to a real-valued equation by keeping only the real part
of $\Rh_z$ and throwing away the uninformative imaginary part. 
%In other words, since the information needed for detecting the OFDM signal is real-valued, we can keep the real part and throw away the uninformative imaginary part.
To avoid introducing extra notations, from now on we assume that (\ref{eq:hRz_vs_Rx}) represents a real-valued matrix equation. 

\myvhskip The problem of sensing OFDM signal can then be restated as testing whether in 2-D linear model (\ref{eq:hRz_vs_Rx}) we have $\mR_x = \tau_0 \mI$ or $\mR_x=\tau_0 \mI + \tau_s \mLambda$ for some 
nonzero unknown parameters $\tau_0$ and $\tau_s$. The first problem in this regard is that (\ref{eq:hRz_vs_Rx}) represents an underdetermined linear system of equations. 
To solve this issue, we first apply the $\vech$ operator to (\ref{eq:hRz_vs_Rx}) and use (\ref{eq:vech}), (\ref{eq:vect}), (\ref{eq:veckron}) and (\ref{eq:Rx_OFDM}) to obtain 
\beq
\rhz= \Hmat_M (\smpMat \otimes \smpMat) (\tau_0 \vect(\mI) + \tau_s \vect(\mLambda) ) + \Hmat_M (\smpMat \otimes \smpMat) \Gmat_N \ww,
\label{eq:intermed1}
\eeq
where $\rhz=\vech(\Rh_z)$ and $\ww=\vech(\mW)$. 
From (\ref{eq:veckron}) and (\ref{eq:veckron2}) it is easy to verify that $(\smpMat \otimes \smpMat) \vect(\mI)=\sum\limits_{i=1}^N (\va_i \otimes \va_i)$ and $(\smpMat \otimes \smpMat) \vect(\mLambda)=\sum\limits_{i=\Td+1}^N (\va_{i-\Td} \otimes \va_i + \va_i \otimes \va_{i-\Td})$ and therefore rewrite (\ref{eq:intermed1}) as
\beqa
\rhz &=& \tau_s \vb_s + \tau_0 \vb_0 + \vv \nonr \\
&=& \mB \thetab + \vv
\label{eq:linmod1}
\eeqa
where $\vb_0 \triangleq \Hmat_M \sum\limits_{i=1}^N (\va_i \otimes \va_i)$ and $\vb_s \triangleq \Hmat_M \sum\limits_{i=\Td+1}^N (\va_{i-\Td} \otimes \va_i + \va_i \otimes \va_{i-\Td})$, $\mB \triangleq [\vb_0~\vb_s]$, $\thetab \triangleq [\tau_0~ \tau_s]^\mytop$, $\vv \triangleq \mK \ww$ is the additive noise vector, and $\mK \triangleq \Hmat_M (\smpMat \otimes \smpMat) \Gmat_N$.

\myvhskip Now, the problem of OFDM signal detection can be expressed as testing whether in the classical linear model (\ref{eq:linmod1}) the second entry of parameter vector $\thetab$ is zero or not. %$[0~1] \thetab = 0$ or not. 
We formulate this in the next section.

We remark here that since in linear regression equation (\ref{eq:linmod1}) the number of unknown parameters is two and the number of equations is $M(M+1)/2$, any $M \ge 2$ is theoretically enough for estimating $\tau_s$ and performing the test.

%In the next section we will formulate the test (\ref{eq:test111}) for classical linear model (\ref{eq:linmod1}).
%%%%%%%%%%%%%%%%%%%%%%%%%%%%%%%%%%%%%%%%%%%%%%%%%%%%%%%%%%%%%%%%%%%%%%%%%%%%%%%%%%%%%%%%%%%%%%%%%%%%%%%%%%%%%%%%%%%%%%%%%%%%%%%%%%%%%%%%%%%%%%%%%%%%%%%%%%%%%%%%%%%%%%%%%%%%%%%%%%%%%%%%%%%%%%%%
%%%%%%%%%%%%%%%%%%%%%%%%%%%%%%%%%%%%%%%%%%%%%%%%%%%%%%%%%%%%%%%%%%%%%%%             SECTION IV            %%%%%%%%%%%%%%%%%%%%%%%%%%%%%%%%%%%%%%%%%%%%%%%%%%%%%%%%%%%%%%%%%%%%%%%%%%%%%%%%%%%%%%
%%%%%%%%%%%%%%%%%%%%%%%%%%%%%%%%%%%%%%%%%%%%%%%%%%%%%%%%%%%%%%%%%%%%%%%%%%%%%%%%%%%%%%%%%%%%%%%%%%%%%%%%%%%%%%%%%%%%%%%%%%%%%%%%%%%%%%%%%%%%%%%%%%%%%%%%%%%%%%%%%%%%%%%%%%%%%%%%%%%%%%%%%%%%%%%%
\section{Approximate Generalized Likelihood Ratio Test}
\label{sec:4}

\myvhskip To derive the GLRT-based detector, we first notice that the noise vector $\vv$ in (\ref{eq:linmod1}) is not white. Therefore the first step is to whiten the noise
by multiplying both sides of (\ref{eq:linmod1}) by $\mSigma_v^{-1/2}$, where \beq\mSigma_v=\mK \mSigma_w \mK^\mytop,\eeq and $\mSigma_w$ is the covariance 
matrix of $\ww=\mbox{vech}(\mW)$. Remark that as proved in Theorem \ref{thm1}, $\mSigma_w$, and therefore $\mSigma_v$, does not expose the same expression under $\cH_0$
and $\cH_1$. 
Hence, to simplify the test, in this section we formulate an GLRT-based detector based on approximating $\mSigma_v$ with a hypothesis-independent covariance matrix which compromises the properties of covariance matrices under the two hypotheses. %The second detector is a data-adaptive one which first estimates $\mSigma_v$ from reduced-rate samples $\{\vsnsig(l)\}_{l=1}^\myL$ and then formulates the detector based on that.

%\subsection{Approximate GLRT-based detector}
%\label{subsecNP1}

We first adopt a hypothesis-independent approximation for the entries of $\mSigma_w$ as follows

\beqa
\Exp(\wij \wpq)=\left\{\begin{array}{ll}
2c & \textrm{if}~(i,j)=(p,q),i-j=0,\\
c & \textrm{if}~(i,j)=(p,q),i-j \neq 0\\
0 & \textrm{otherwise},\end{array}
\right.
\label{eq:cov_w_approx}
\eeqa
where $c$ is an unknown constant. Remark that $\Exp(\wij \wpq)$ in (\ref{eq:cov_w_approx}) shares properties with both $\Exp(\wij \wpq|\cH_0)$ in (\ref{eq:covar_h0}) and $\Exp(\wij \wpq|\cH_1)$ in (\ref{eq:var_h1}): like (\ref{eq:covar_h0}) it is zero for $(i,j)\neq(p,q)$, but then for $(i,j)=(p,q)$ it is like (\ref{eq:var_h1}) with $\tau_s \ll \tau_0$. Equation (\ref{eq:cov_w_approx}) then implies that $\mSigma_w$ has the following form
\beq
\mSigma_w = c\mDelta,
\label{eq:noisecov}
\eeq
where $\mDelta$ is an $\frac{N(N+1)}{2}$ diagonal matrix with diagonal elements in locations $1$, $1+(N)$, $1+(N)+(N-1)$, $1+(N)+(N-1)+(N-2)$, ... being equal to $2$ and the rest are equal to $1$. In fact, the diagonal
elements having value 2 correspond to the first case of (\ref{eq:cov_w_approx}) and the rest correspond to its second case. The third case of (\ref{eq:cov_w_approx}) implies that $\mSigma_w$ is diagonal. 

\myvhskip Now, multiplying both sides of (\ref{eq:linmod1}) by $\mGamma=(\mK \mDelta \mK^\mytop)^{-1/2}$ yields:
\beqa
\rrz = \mBB \thetab + \vvv,
\label{eq:linmod3}
\eeqa
where $\rrz \triangleq \mGamma \rhz$, $\mBB \triangleq \mGamma \mB$, and $\vvv \triangleq \mGamma \vv$. 
Based on this, the problem of OFDM detection can be expressed as doing hypothesis testing problem (\ref{eq:test111}) for linear system (\ref{eq:linmod3}) in which the additive noise has distribution $\vvv \sim \cN(0,c\mI)$ with some unknown variance $c$. The GLRT-based detector for this problem 
can then be written as \citep[Theorem 9.1]{kay_volII} 
%The theorem states that the test statistic for doing hypothesis (\ref{eq:test111}) is  
\beqa
T(\rrz) \triangleq \frac{(\Nef-2) (\mSS \hat{\thetab}_{1})^\mytop [\mSS(\mBB^\mytop \mBB)^{-1} \mSS^\mytop]^{-1}(\mSS \hat{\thetab}_{1})}{\rrz^\mytop\Big(\mI-\mBB (\mBB^\mytop \mBB)^{-1} \mBB^\mytop\Big)\rrz} %\nonr \\
\overset{\cH_1}{\underset{\cH_0}{\gtrless}} \gamma^\prime
\label{eq:test_approximate} %\\
%&=& (\Nef-1)\frac{(\vbbb_s^\mytop \rrrz)\|\vbbb_s\|_2^2}{\|\vbbb_s\|_2^2 \|\rrrz\|_2^2-(\vbbb_s^\mytop \rrrz)^2}, 
%&=& \frac{(N-1)<\rrrz,\vbbb><\vbbb,\vbbb>}{<\rrrz,\rrrz><\vbbb,\vbbb>-<\rrrz,\vbbb><\rrrz,\vbbb>}
\eeqa
where $\Nef\triangleq\frac{M(M+1)}{2}$ denotes the number of distinct equations, $\mSS\triangleq[0~1]$, and $\hat{\thetab}_{1}=(\mBB^\mytop \mBB)^{-1} \mBB^\mytop \rrz$ is the maximum likelihood estimate of $\thetab$ under $\cH_1$. 
%Then, we will decide in favor of $\cH_1$ (an OFDM signal exists) if
%\beq
%T(\rrrz) > \gamma^\prime,
%\eeq
%and in favor of $\cH_0$ (no OFDM signal) otherwise. 
Furthermore, the probability of false alarm ($\pfa$) and probability of detection ($\pdet$) of the approximate GLRT-based detector are given by 
\beqa
\pfa=Q_{F_{1,\Nef-2}}(\gamma^\prime), \label{eq:NP_pfa}\\
\pdet=Q_{F_{1,\Nef-2}(\lambda)}(\gamma^\prime), \label{eq:NP_pd} 
\eeqa
where the noncentrality parameter takes the form
\beq
\lambda = \frac{(\mSS \thetab_1)^\mytop [\mSS(\mBB^\mytop \mBB)^{-1}\mSS^\mytop]^{-1}(\mSS \thetab_1)}{c},
\label{ncparam}
\eeq
where $\thetab_1$ is the true value of $\thetab$ under $\cH_1$.

\section{Extension to Frequency-Selective Fading Channels}
\label{sec:multipath}

In this section we extend the results obtained in the previous sections to the case of wideband frequency-selective channel.
Assume that the multipath channel between the PU transmitter and the SU receiver has the model $H(z)=h_0+h_1 z^{-1}+\ldots+h_\myL z^{-\myL}$. Putting the tap weights in $(\myL+1)\times 1$
vector $\vh\triangleq[h_0,h_1,\ldots,h_{\myL}]^\mytop$ and denoting the $k$-th OFDM block of length $\Tsym=N$  as $\vs_k \triangleq [s_k(1),s_k(2),\ldots,s_k(N)]$ where $s_k(n)\triangleq s[(k-1)N+n]$, we have
\beq
\vx_k = \mS_k \vh + \vn,~k=1,2,\ldots,\Nblk, 
\label{eq:multipath_model}
\eeq
where $\mS_k$ is a Toeplitz $N \times (L+1)$ matrix defined as %in (\ref{eq:Sk}).
%\begin{figure*}[t!]
\beqa
\mS_k \triangleq \left[ \begin{array}{ccccc} 
s_k(1) & s_{k-1}(N) & s_{k-1}(N-1) & \ldots & s_{k-1}(N-L+1) \\
s_k(2) & s_{k}(1) & s_{k-1}(N) & \ldots & s_{k-1}(N-L+2) \\
s_k(3) & s_{k}(2) & s_{k}(1) & \ldots & s_{k-1}(N-L+3) \\
\ldots & \ldots & \ldots & \ldots & \ldots \\
s_k(N) & s_{k}(N-1) & s_{k}(N-2) & \ldots & s_{k}(N-L) \\
\end{array} \right].
\label{eq:Sk}
\eeqa 
%\end{figure*}
Denoting the $i$-th column of $\mS_k$ by $\vs_{k,i}$ we can write
\beq
\vs_{k,i} =  \mJ_d^i \vs_k + \mJ_u^{N-i} \vs_{k-1},
\label{eq:vsk}
\eeq
where $\mJ_d$ and $\mJ_u$ are respectively down-shift matrix and up-shift matrix whose $(i,j)$-th elements are $[\mJ_d]_{i,j}\triangleq\delta_{i,j+1}$ and  $[\mJ_u]_{i,j}\triangleq\delta_{i+1,j}$ where $\delta_{i,j}$ denotes the Kronecker delta. Remark that $\mJ_d=\mJ_u^T$.

From \citep[Theorem 21.6]{seber08} and since $\Exp(\mS_k)=\mZero_{N,\myL+1}$ we can next write
\beqa
\mR_x &=& \Exp(\mS_k \vh \vh^H \mS_k^H) + \sigma_0^2 \mI \nonr \\
&=& \sum_{i=0}^\myL \sum_{j=0}^\myL h_i h_{j}^\ast \mbox{Cov} (\vs_{k,i},\vs_{k,j}) + \sigma_0^2 \mI.
\label{eq:covmat}
\eeqa
From (\ref{eq:vsk}), $\mbox{Cov} (\vs_{k,i},\vs_{k,j})$ can be computed as
\beq
\mbox{Cov} (\vs_{k,i},\vs_{k,j}) = \sigma_s^2 \Big(\mJ_d^i (\mLambda+\mI) \mJ_u^j + \mJ_u^{N-i} (\mLambda+\mI) \mJ_d^{N-j} \Big)
\label{eq:covij}
\eeq
Denoting $\vc_{i,j} \triangleq \vect\Big(\mJ_d^i (\mLambda+\mI) \mJ_u^j + \mJ_u^{N-i} (\mLambda+\mI) \mJ_d^{N-j} \Big)$, we can write
%Denoting $\vc_{i,j} \triangleq \frac{1}{\sigma_s^2}\vect(\mbox{Cov} (\vs_{k,i},\vs_{k,j}))$ we can write.......$\{(j(\myL+1)+i+1)\}_{(i,j)=(0,0)}^{(L,L)}$
\beqa
\vect(\mR_x)&=& \sum_{i=0}^\myL \sum_{j=0}^\myL h_i h_{j}^\ast \vc_{i,j} + \sigma_0^2 \vect(\mI) \nonr \\
&=& [\mC_s,\vect(\mI)] \left[\begin{array}{c} \tilde{\vh} \\ \sigma_0^2 \end{array}\right],
\label{eq:multipath_Rx}
\eeqa 
where $\mC_s$ is defined as the $N^2 \times (L+1)^2$ matrix whose $(j(\myL+1)+i+1)$-th column is $\vc_{i,j}$ (remark that $0 \le i,j \le L$) and $\tilde{\vh}$ is defined as the vector whose $(j(\myL+1)+i+1)$-th entry is $\sigma_s^2 h_i h_j^\ast$.

Putting (\ref{eq:multipath_Rx}) in (\ref{eq:hRz_vs_Rx}) and vectorizing the resulting equation yields
\beq
\rhz = \mB_m \thetab_m + \vv
%\rhz = \Hmat_M (\smpMat^\ast \otimes \smpMat) [\mC_s,\vect(\mI)]  + \Hmat_M (\smpMat \otimes \smpMat) \Gmat_N \ww,
\label{eq:mp_intermed1}
\eeq
where $\mB_m \triangleq \Hmat_M (\smpMat \otimes \smpMat) [\mC_s,\vect(\mI)] $, $\thetab_m \triangleq [ \tilde{\vh}^T,\sigma_0^2]^T$, and $\vv \triangleq \Hmat_M (\smpMat \otimes \smpMat) \Gmat_N \vech(\mW)$.
Then the problem of OFDM detection can be expressed as performing the following test for the linear model of (\ref{eq:mp_intermed1})
\beqa
\left\{ \begin{array}{l} \cH_0: \mSS_m \thetab_m = \mZero, \\ 
\cH_1: \mSS_m \thetab_m \neq \mZero, \end{array} \right.
\label{eq:test_freqsel}
\eeqa
where $\mSS_m =[\mI_{(L+1)^2},\mZero_{(L+1)^2,1}]$. 
The rest of the test formulation is similar to the Gaussian case discussed in \ref{sec:4}.

\section{Simulation Examples}
\label{sec:5}

\myvhskip In this section we study the performance of the proposed sub-Nyquist OFDM detector by simulation examples.
For simplicity, we consider an OFDM system with IFFT size $32$ which means $\Td=32$. The cyclic prefix length is chosen as $\Tcp=\Td/4=8$ and 
subcarrier symbols are assumed to be drawn from a 16-QAM constellation with unit energy. The elements of the measurement matrix $\mA$ are drawn from $\mathcal{N}(0,1)$ and its columns are normalized to have unit norm. We emphasize that these chosen values are just examples for carrying out numerical simulations, and are not as such related to the fundamentals of the derived detectors in any way. 

\myvhskip Before proceeding with the simulation examples, we briefly discuss the parameters which affect the performance based on the
formulation of the problem given in previous sections. We first recall that the linear equation used for detecting the OFDM signal
is (\ref{eq:hRz_vs_Rx}). Therefore the parameters which affect the detection are those affecting the solution of this 2-D linear regression problem.
The first parameter is the number of independent equations in (\ref{eq:hRz_vs_Rx}), i.e., $M(M+1)/2$. For a fixed Nyquist frame size $N$, this parameter is uniquely  
specified by the compression ratio $\rho$. The second and third parameters are those which affect the power of finite sample error $\mW$. Formulas (\ref{eq:covar_h0})-(\ref{eq:var_h1}) clearly show
that, for a given signal power $\sigma_s^2$, this is determined by $\Nblk$; the number of blocks taken for computing $\mR_z$, and $\sigma_n^2$; the 
variance of noise, or equivalently, signal to noise ratio. The first two experiments in this section are devised based on the above observations. Besides, the exactness of the approximation we adopted in Section \ref{sec:4} will be studied by simulation results.
As mentioned in Section \ref{sec:intro} there is no method in the literature specifically designed for single-band OFDM detection from sub-Nyquist samples. Therefore, to compare our method with some existing methods, we choose the method proposed in \cite[Section IV]{tian12} which recovers the PSD of a general stationary signal from sub-Nyquist samples.

The third experiment studies the performance of the method over frequency-selective channels.

\subsection{Influence of the compression ratio} 
\begin{figure}[t!]
                \centering
                \includegraphics[width=1\textwidth,height=0.65\textwidth]{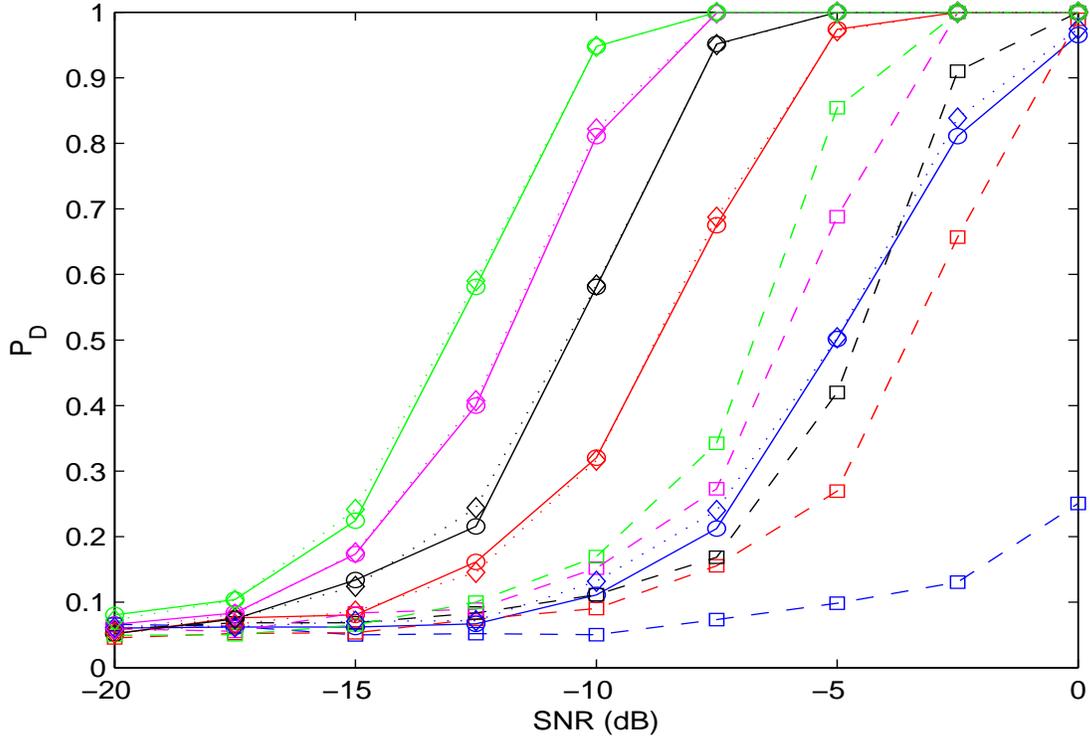}
	 \caption{Probability of Detection versus SNR for various compression ratios $\rho$. The number of blocks is $\Nblk=100$ and the probability of false-alarm is $\pfa=0.05$. The solid curves with circle markers correspond to the case where the threshold has been determined based on the true $P_{FA}$ computed from the simulated data. The dotted curves with diamond markers correspond to the case where the threshold has been determined from formula (\ref{eq:NP_pfa}). The dashed curves with square markers correspond to the PSD recovery method introduced in \cite{tian12}. Color conventions are as follows. Blue: $\rho=0.2$, red: $\rho=0.4$, black: $\rho=0.6$, magenta: $\rho=0.8$, green: $\rho=1$. As it can be seen the proposed algorithm clearly outperforms the method introduced in \cite{tian12}. Furthermore, the very small distance between the solid curves with circle markers and the dotted curves with diamond markers shows the exactness of the approximation we adopted.}
	\label{fig:1-1}
	\vskip-0.5cm\end{figure}

In the first experiment we study the effect of compression ratio on detection performance. For this experiment, the detector deploys $\Nblk=100$ OFDM symbols.
Figure \ref{fig:1-1} illustrates the
probability of detection of the proposed method as a function of signal-to-noise ratio (SNR) for different compression ratios $\rho \in \{0.2,0.4,0.6,0.8,1\}$
when the threshold is set to obtain $\pfa=0.05$. To check also the exactness of the approximation adopted in Section \ref{sec:4}, we compute the threshold $\gamma^\prime$ both from the true $P_{FA}$ calculated from the simulated data in the absence of OFDM signal (solid lines with circle markers) and from formula (\ref{eq:NP_pfa}) (dotted lines with diamond marker). The results of the PSD recovery method introduced in \cite{tian12} are shown by dashed lines with square markers. For this method the threshold has been computed from the true $P_{FA}$.
 
\begin{figure}[t!]
                \centering
                \includegraphics[width=1\textwidth,height=0.65\textwidth]{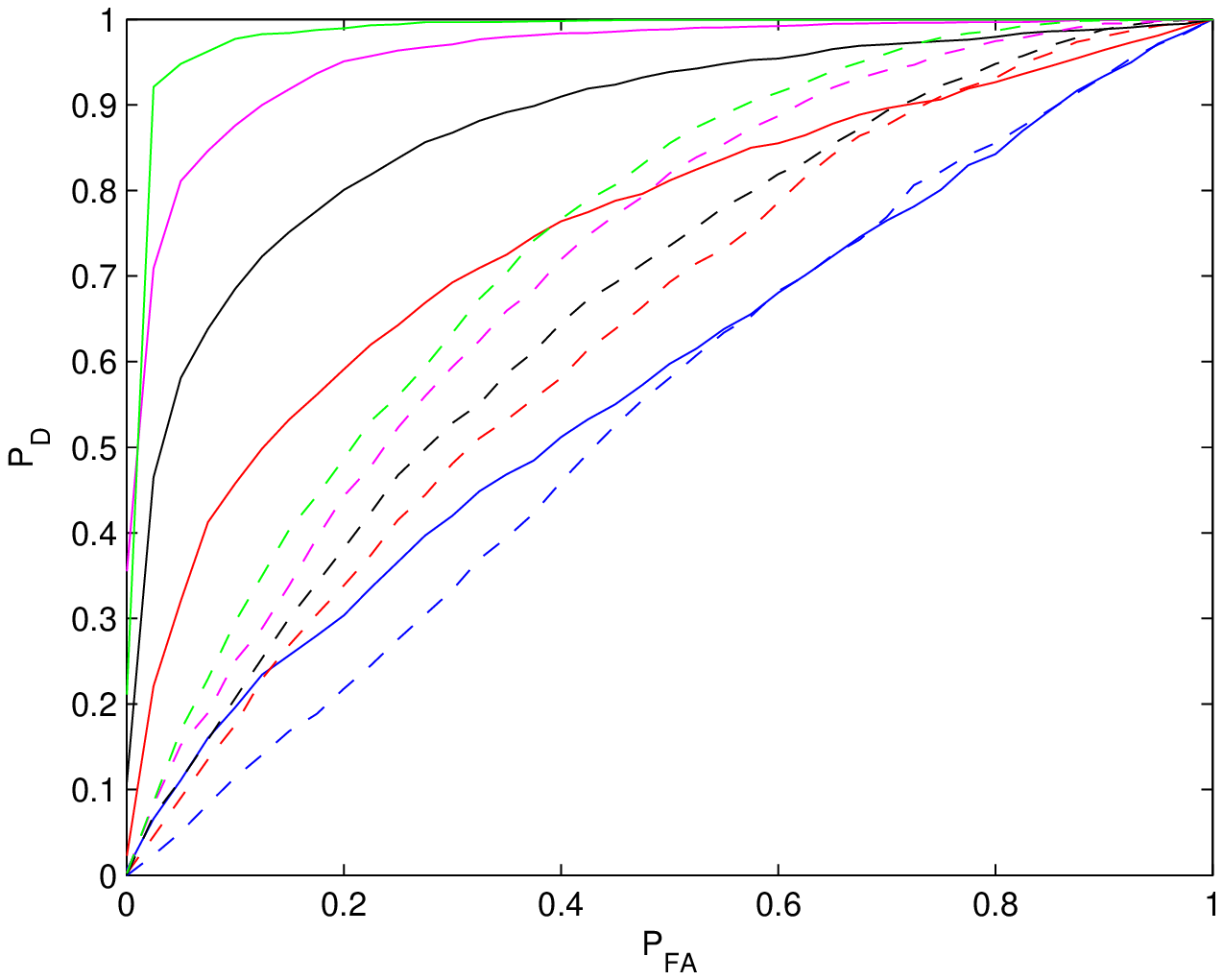}
	 \caption{Probability of Detection versus SNR for various compression ratios $\rho$. The number of blocks is $\Nblk=100$ and the probability of false-alarm is $\pfa=0.05$. The solid curves correspond to the case where the threshold has been determined based on the true $P_{FA}$ computed from the simulated data. The dashed curves correspond to the PSD recovery method introduced in \cite{tian12}. The color conventions are as in Figure \ref{fig:1-1}. As it can be seen the proposed algorithm clearly outperforms the method introduced in \cite{tian12}. }
	\label{fig:1-2}
	\vskip-0.5cm\end{figure}

The first observation from \ref{fig:1-1} is that, as we expect, the performance of the proposed method enhances with increasing the compression ratio. Interestingly, with heavy compression ratio of 0.2, the detection probability is still above $95\%$ at an SNR of $0$ dB. 
Notice that since the number of independent covariance-based equations is of order $\mathcal{O}(M^2)$, a compression ratio of $\rho \triangleq M/N=0.2$ implies that the {\it effective} compression ratio (in covariance domain) is only about $0.2^2=0.04$ or $4\%$. 
Also, as it can be observed, the proposed method provides a much better performance compared to the general PSD recovery method of \cite[Section IV]{tian12}. 
This can be seen also from Figure \ref{fig:1-2} where the Receiver Operating Characteristic (ROC) curves of both methods have been depicted. 

Furthermore, the unnoticeable difference between the curves with threshold computed from true $P_{FA}$ and the ones with threshold computed from (\ref{eq:NP_pfa}) in Figure \ref{fig:1-1} verifies the exactness of the approximation adopted in Section \ref{sec:4}.

%Remark that the effective
%compression ratio here is $\frac{M(M+1)/2}{N(N+1)/2}$ which is the number of unique equations divided by the number of unique 
%unknowns in ().
\subsection{Influence of the number of blocks} 
\begin{figure}[t!]
                \centering
                \includegraphics[width=1\textwidth,height=0.65\textwidth]{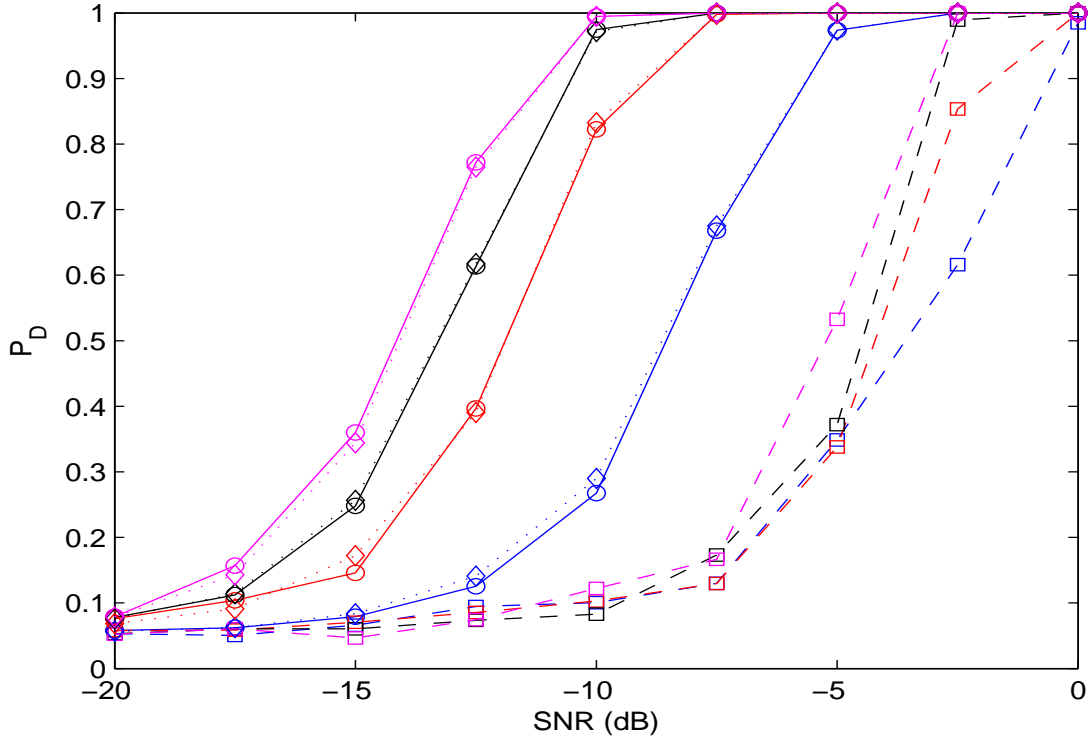}
	 \caption{Probability of Detection versus SNR for various number of blocks $\Nblk$. The compression ratio is $\rho=0.4$ and the probability of false-alarm is $\pfa=0.05$. The solid curves with circle markers correspond to the case where the threshold has been determined based on the true $P_{FA}$ computed from the simulated data. The dotted curves with diamond markers correspond to the case where the threshold has been determined from formula (\ref{eq:NP_pfa}). The dashed curves with square markers correspond to the PSD recovery method introduced in \cite{tian12}. Color conventions are as follows. Blue: $\Nblk=100$, red: $\Nblk=400$, black: $\Nblk=700$, magenta: $\Nblk=1000$. As it can be seen the proposed algorithm clearly outperforms the method introduced in \cite{tian12}. Furthermore, the very small distance between the solid curves with circle markers and the dotted curves with diamond markers shows the exactness of the approximation we adopted.}
	\label{fig:2-1}        
\vskip-0.5cm\end{figure}

\begin{figure}[t!]
                \centering
                \includegraphics[width=1\textwidth,height=0.65\textwidth]{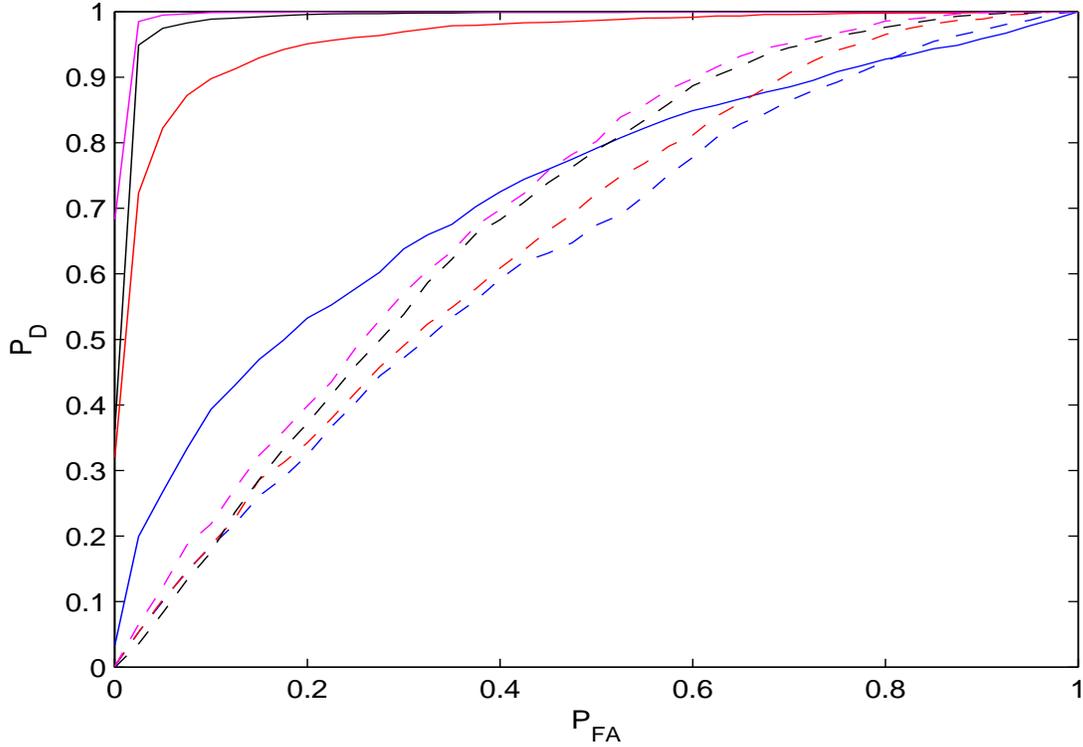}
	 \caption{Probability of Detection versus SNR for various number of blocks $\Nblk$. The compression ratio is $\rho=0.4$ and the probability of false-alarm is $\pfa=0.05$. The solid curves correspond to the case where the threshold has been determined based on the true $P_{FA}$ computed from the simulated data. The dashed curves correspond to the PSD recovery method introduced in \cite{tian12}. The color conventions are as in Figure \ref{fig:2-1}. As it can be seen the proposed algorithm clearly outperforms the method introduced in \cite{tian12}. }
	\label{fig:2-2}
	\vskip-0.5cm\end{figure}

In the second experiment we study the effect of number of blocks taken for computing the covariance matrix, $\Nblk$, on detection performance.
As it can be seen from Figure \ref{fig:2-1} increasing $\Nblk$ improves the performance of the detector. In fact, in asymptotic case when $\Nblk\rightarrow\infty$
we have $\bar{\mR}_x=\frac{1}{\Nblk} \sum_{l=1}^\Nblk(\vx(l) \vx(l)^H)=\Exp(\vx \vx^H)=\mR_x$ which means $\mW=\mZero$.
Similar to the previous experiment, it can be also observed that the performance of the proposed method shows significance improvement over the PSD recovery method of \cite{tian12}.
This can be seen also from Fiure \ref{fig:2-2} where the ROC curves of both methods have been illustrated.

Again, the unnoticeable difference between the solid lines (true $P_{FA}$) and their corresponding dotted lines (target $P_{FA}$) in Figure \ref{fig:2-1} confirms the exactness of the approximation we used in Section \ref{sec:4}.

\subsection{Performance of the proposed method over frequency-selective channels}
\begin{figure}[t!]
        \centering
        %\begin{subfigure}[t]%{0.5\textwidth}
                \centering
                \includegraphics[width=1\textwidth,height=0.65\textwidth]{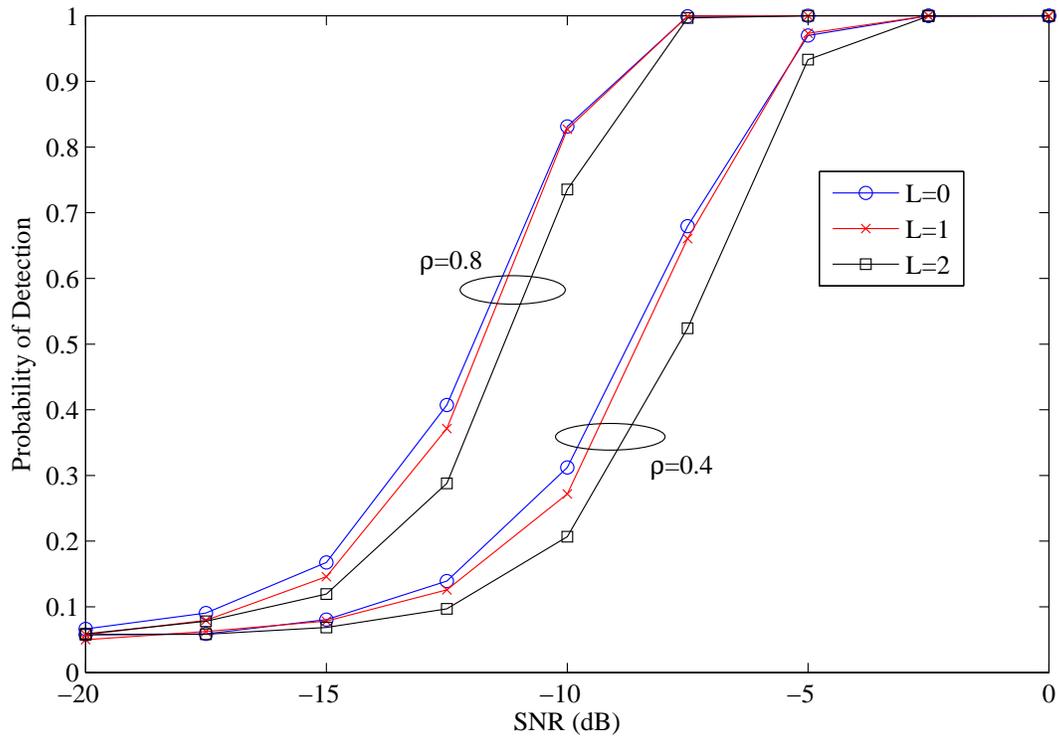}
	\caption{Effect of frequency-selectivity of the channels on the performance of the proposed method. The channel length is $L+1$ and $\rho$ is the compression ratio.}
	\label{fig:3} 
\end{figure}

The third experiment studies the effect of frequency-selective channels on the performance of the proposed method. 
The method is applied on three different channel lengths with identical energy; i.e. $\sum_{l=0}^\myL |h_l|^2$ is identical for all $\myL=0,1,2$.
The OFDM signal characteristics as well as the probability of false-alarm are set to the same values as in the two previous examples. As it can be observed from Figure \ref{fig:3} the frequency selectivity
slightly deteriorates the performance but reliable sensing can still be achieved even at low SNRs.

%%%%%%%%%%%%%%%%%%%%%%%%%%%%%%%%%%%%%%%%%%%%%%%%%%%%%%%%%%%%%%%%%%%%%%%%%%%%%%%%%%%%%%%%%%%%%%%%%%%%%%%%%%%%%%%%%%%%%%%%%%%%%%%%%%%%%%%%%%%%%%%%%%%%%%%%%%%%%%%%%%%%%%%%%%%%%%%%%%%%%%%%%%%%%%%%
%%%%%%%%%%%%%%%%%%%%%%%%%%%%%%%%%%%%%%%%%%%%%%%%%%%%%%%%%%%%%%%%%%%%%%%             SECTION VI            %%%%%%%%%%%%%%%%%%%%%%%%%%%%%%%%%%%%%%%%%%%%%%%%%%%%%%%%%%%%%%%%%%%%%%%%%%%%%%%%%%%%%%
%%%%%%%%%%%%%%%%%%%%%%%%%%%%%%%%%%%%%%%%%%%%%%%%%%%%%%%%%%%%%%%%%%%%%%%%%%%%%%%%%%%%%%%%%%%%%%%%%%%%%%%%%%%%%%%%%%%%%%%%%%%%%%%%%%%%%%%%%%%%%%%%%%%%%%%%%%%%%%%%%%%%%%%%%%%%%%%%%%%%%%%%%%%%%%%% 
\section{Concluding Remarks}
\label{sec:conclusion}
A method for sensing OFDM signals from sub-Nyquist samples was proposed. The proposed method exploits the
unique characteristics of the covariance matrix of OFDM signal to perform the spectrum sensing task. Based on the statistical properties of the 
covariance matrix we developed an approximate GLRT-based detector. The proposed method was also extended to the case of frequency-selective channels. The simulation results verify the theoretical observations. 
The results also illustrate that highly efficient sensing can still be obtained, in terms of probability of detection and false alarm, despite of low compression ratios and low SNRs. This can open up new possibilities for sensing devices with low-cost analog hardware and A/D interface deploying sub-Nyquist observations.

\section*{Acknowledgments}
The work of S. A. Razavi and M. Valkama was supported by the Academy of Finland under the  project $\#$251138 ``Digitally-Enhanced RF for Cognitive Radio Devices'' and the Finnish Funding Agency for Technology and Innovation (Tekes), under the project ``Enabling Methods for Dynamic Spectrum Access and Cognitive Radio (ENCOR)''.

%%%%%%%%%%%%%%%%%%%%%%%%%%%%%%%%%%%%%%%%%%%%%%%%%%%%%%%%%%%%%%%%%%%%%%%%%%%%%%%%%%%%%%%%%%%%%%%%%%%%%%%%%%%%%%%%%%%%%%%%%%%%%%%%%%%%%%%%%%%%%%%%%%%%%%%%%%%%%%%%%%%%%%%%%%%%%%%%%%%%%%%%%%%%%%%%
%%%%%%%%%%%%%%%%%%%%%%%%%%%%%%%%%%%%%%%%%%%%%%%%%%%%%%%%%%%%%%%%%%%%%%%              APPENDIX             %%%%%%%%%%%%%%%%%%%%%%%%%%%%%%%%%%%%%%%%%%%%%%%%%%%%%%%%%%%%%%%%%%%%%%%%%%%%%%%%%%%%%%
%%%%%%%%%%%%%%%%%%%%%%%%%%%%%%%%%%%%%%%%%%%%%%%%%%%%%%%%%%%%%%%%%%%%%%%%%%%%%%%%%%%%%%%%%%%%%%%%%%%%%%%%%%%%%%%%%%%%%%%%%%%%%%%%%%%%%%%%%%%%%%%%%%%%%%%%%%%%%%%%%%%%%%%%%%%%%%%%%%%%%%%%%%%%%%%%
\section*{Appendix: \bf Proof of Theorem \ref{thm1}}
%The proofs for (\ref{eq:var_h0}) and the second and third cases of (\ref{eq:var_h0}) have been given in \citep{axell11}. 
The proof of (\ref{eq:covar_h0}) is trivial. Here we only prove (\ref{eq:var_h1}).
Let us define $r_{i,j}\triangleq\mathcal{\mbox{Real}}\Big([\Rh_x]_{i,j}\Big)$, and denote the real part of $x_i$ by $\xrl_i$ and its imaginary part by $\xiy_i$.
Since 
\beqa r_{i,j}&=&\mathcal{\mbox{Real}}(\frac{1}{\Nblk} \sum_{l=1}^\Nblk(x_i(l) x_j^*(l)))\nonr \\ &=&\frac{1}{\Nblk} \sum_{l=1}^\Nblk(\xrl_i(l) \xrl_j(l)+\xiy_i(l) \xiy_j(l)),\nonr\eeqa 
we can write
\beqa \begin{array}{lll}
\rij \rpq &=& \frac{1}{\Nblk^2} \sum\limits_{l=1}^{\Nblk} \sum\limits_{l^\prime=1}^{\Nblk} \Big([\xrl_i(l) \xrl_j(l)+\xiy_i(l) \xiy_j(l)] \\
&&[\xrl_p(l^\prime) \xrl_q(l^\prime)+\xiy_p(l^\prime) \xiy_q(l^\prime)]\Big) \\
&=& \Tone+\Ttwo, \end{array} \label{eq:app1}\eeqa
where $\Tone$ and $\Ttwo$ are defined as follows
\beqa \begin{array}{lll}
\Tone &\triangleq& \frac{1}{\Nblk^2} \sum\limits_{l=1}^{\Nblk} \sum\limits_{\substack {l^\prime=1 \\ \l^\prime \neq l}}^{\Nblk} \Big([\xrl_i(l) \xrl_j(l)+\xiy_i(l) \xiy_j(l)] \\
&&[\xrl_p(l^\prime) \xrl_q(l^\prime)+\xiy_p(l^\prime) \xiy_q(l^\prime)]\Big), \\%\end{array}\\
%\eeqa
%\beqa \begin{array}{lll}
\Ttwo &\triangleq& \frac{1}{\Nblk^2} \sum\limits_{l=1}^{\Nblk} \Big([\xrl_i(l) \xrl_j(l)+\xiy_i(l) \xiy_j(l)] \\
&&[\xrl_p(l) \xrl_q(l)+\xiy_p(l) \xiy_q(l)]\Big) \end{array} \nonr
\eeqa
\myvhskip The reason for partitioning (\ref{eq:app1}) to $\Tone$ and $\Ttwo$ is that in $\Tone$ the two multiplicative terms inside the double summation are independent since they belong to
different frames $l$ and $l^\prime$, while in $\Ttwo$ they belong to the same frame and depending on indices $i$, $j$, $p$, and $q$ might be dependent.
Due to independence of multiplicative terms, $\Exp(\Tone|\cH_1)$ is easy to compute as
\beqa
\Exp(\Tone|\cH_1) = \left \{ \begin{array}{ll} 
\frac{\Nblk-1}{\Nblk} \tau_0^2 & \mbox{if}~(i-j,p-q)=(0,0) \\[.25em] 
\frac{\Nblk-1}{\Nblk} \tau_0 \tau_s & \mbox{if}~(i-j,p-q)\in\{(0,\Td),(\Td,0)\}\\[.25em]
\frac{\Nblk-1}{\Nblk} \tau_s^2 & \mbox{if}~(i-j,p-q)=(\Td,\Td)\\[.25em]
0 & \textrm{otherwise}. %~~~~~~~~~~~~~~~~~~~~~~~~~~(\arabic{equation})\label{eq:t1} 
 \end{array}\right.
\label{eq:t1}\eeqa
%\addtocounter{equation}{1}
Furthermore, after some more scrutinized manipulations, $\Exp(\Ttwo|\cH_1)$ can be expressed as
\beqa
\Exp(\Ttwo|\cH_1) = \left \{ \begin{array}{ll} 
\frac{2}{\Nblk} \tau_0^2 & \mbox{if}~(i,j)=(p,q), i-j=0, \\[.25em] 
\frac{1}{2\Nblk} (\tau_0^2 + 3\tau_s^2) & \mbox{if}~(i,j)=(p,q),i-j=\Td,\\[.25em]
\frac{1}{2\Nblk} \tau_0^2 & \mbox{if}~(i,j)=(p,q),i-j \notin \{0,\Td\},\\[.25em]
\frac{2}{\Nblk} \tau_0 \tau_s & \mbox{if}~i=j=p=q+\Td,\\[.25em]
\frac{1}{\Nblk} \tau_0^2 & \mbox{if}~(i-j,p-q)=(0,0),i\neq p\\[.25em]
\frac{1}{\Nblk} \tau_s^2 & \mbox{if}~(i-j,p-q)=(\Td,\Td),i\neq p\\[.25em]
\frac{1}{\Nblk} \tau_0 \tau_s & \mbox{if}~(i-j,p-q)\in\{(0,\Td),(\Td,0)\},\\ & ~~~~~~~~~~~~~~~~~~~~~~i\neq p\\[.25em]
0 & \textrm{otherwise}. %~~~~~~~~~~~~~~~~~~~~~~(\arabic{equation})\label{eq:t2}
%0 & \mbox{otherwise}.  
\end{array}\right. \label{eq:t2} \eeqa
%\beq\label{eq:t2}\eeq
%\addtocounter{equation}{1}
Summing (\ref{eq:t1}) and (\ref{eq:t2}) will yield then
\begin{eqnarray}
\Exp(\rij \rpq|\cH_1) = \left \{ \begin{array}{ll} 
\tau_0^2 + \frac{\tau_0^2}{\Nblk}  & \mbox{if}~(i,j)=(p,q), i-j=0, \\[.2em] 
\tau_s^2 + \frac{\tau_0^2+\tau_s^2}{2\Nblk} & \mbox{if}~(i,j)=(p,q),i-j=\Td,\\[.2em]
\frac{\tau_0^2}{2\Nblk}  & \mbox{if}~(i,j)=(p,q),i-j \notin \{0,\Td\},\\[.2em]
\tau_0 \tau_s + \frac{\tau_0 \tau_s}{\Nblk}  & \mbox{if}~i=j=p=q+\Td,\\[.2em]
\tau_0^2 & \mbox{if}~(i-j,p-q)=(0,0),i\neq p\\[.2em]
\tau_s^2 & \mbox{if}~(i-j,p-q)=(\Td,\Td),i\neq p\\[.2em]
\tau_0 \tau_s & \mbox{if}~(i-j,p-q)\in\{(0,\Td),\\ & ~~~~~~~~~~~~~~~~~~~~,(\Td,0)\},i\neq p\\[.2em]
0 & \textrm{otherwise}. %~~~~~~~~~~~~~~~~~~~~(\arabic{equation})\label{eq:t12}
%0 & \mbox{otherwise}.  
\end{array}\right. \label{eq:t12} \end{eqnarray}
%\beq\label{eq:t12}\eeq
Subtracting $\Exp(\rij|\cH_1) \Exp(\rpq|\cH_1)$ from (\ref{eq:t12}) and remarking that $\wij$ and $\rij$ have identical variances, (\ref{eq:var_h1}) is concluded.

% -------------------------------------------------------------------------
%\bibliographystyle{plainnat}
%%\bibliographystyle{elsarticle-num}
%\bibliography{itinsp}
%%\end{thebibliography}

%\newpage
%\begin{figure}[t!]
                %\centering
                %\includegraphics[scale=0.5]{diagram2.pdf}
	 %\caption{Block diagram of the system we use for deciding on occupancy of the band.}
	%\label{fig:diag}
	%\vskip0.5cm\end{figure} 
%
%
%\clearpage
\newpage
	
%\begin{figure}
%        \centering
%        \begin{subfigure}[b]{\textwidth}
%        		\includegraphics[width=0.8\textwidth,height=0.6\textwidth]{fig1_4.eps}
%       	\end{subfigure}
%        \begin{subfigure}[b]{\textwidth}
%        		\includegraphics[width=0.8\textwidth,height=0.6\textwidth]{fig1_2.eps}
%        \end{subfigure}
%	 \caption{Probability of Detection versus SNR for various compression ratios $\rho$. The number of blocks is $\Nblk=100$ and the probability of false-alarm is $\pfa=0.05$.}
%	\label{fig:1-1} 
%	\end{figure} 
        %\end{subfigure}\\%
        ~ %add desired spacing between images, e. g. ~, \quad, \qquad etc.
        %\begin{subfigure}[t]%{0.5\textwidth}

 \newpage

\newpage
\end{document}